\documentclass[showpacs,preprint,pra]{revtex4}
\def\rhoo{\mbox{\boldmath{$\rho$}}}
\usepackage{dcolumn,longtable}
\begin{document}

\title{
Long-range interactions between a He($2\,^3\!S$) atom and a  He($2\,^3\!P$) atom for like
isotopes}

\author{J.-Y. Zhang$^1$, Z.-C. Yan$^1$, D. Vrinceanu$^{2}$, J. F. Babb$^3$, and H. R.
Sadeghpour$^3$ } \affiliation{$^1$Department of Physics, University of New Brunswick,
Fredericton, New Brunswick, Canada E3B 5A3} \affiliation{$^2$T-4 Group, Los Alamos National
Laboratory, Los Alamos, New Mexico 87545, USA} \affiliation{$^3$ITAMP, Harvard-Smithsonian
Center for Astrophysics, Cambridge, Massachusetts 02138, USA}
\date{\today}

\begin{abstract}
  For the interactions between a He($2\,^3\!S$) atom and a
  He($2\,^3\!P$) atom for like isotopes, we report perturbation
  theoretic calculations using accurate variational wave functions in
  Hylleraas coordinates of the coefficients determining the potential
  energies at large internuclear separations.  We evaluate the
  coefficient $C_{3}$ of the first order resonant dipole-dipole energy
  and the van der Waals coefficients $C_{6}$, $C_{8}$, and $C_{10}$
  for the second order energies arising from the mutual perturbations
  of instantaneous electric dipole, quadrupole, and octupole
  interactions.  We also evaluate the coefficient $C_{9}$ of the
  leading contribution to the third order energy. We establish
  definitive values including treatment of the finite nuclear mass for
  the ${}^3$He($2\,^3\!S$)--${}^3$He($2\,^3\!P$) and
  ${}^4$He($2\,^3\!S$)--${}^4$He($2\,^3\!P$) interactions.
\end{abstract}
 \maketitle


Recently, there has been considerable interest in the study of helium dimers and ultracold
helium collisions associated with metastable helium
atoms~\cite{cohen,cohen2,koelemeij,kim,venleo,kim05,leonard05,sadeg1,sadeg3,flann,singer}.
Molecular lines of the helium dimers associated with the He$(2\,^3\!S)$--He$(2\,^3\!P)$
asymptotes have been produced by photoassociation of spin-polarized metastable helium atoms
He$(2\,^3\!S_1)$.  In the purely long-range $0^{+}_u$ potential well (with a minimum
internuclear distance reaching $150$ bohr) associated with the
He$(2\,^3\!S_1)$--He$(2\,^3\!P_0)$ asymptote, bound states of the helium
dimer~\cite{cohen,cohen2,venleo} can be used in measurements of the $s$-wave scattering
length for collisions of two $^4\!\,$He$(2\,^3\!S_1)$ atoms~\cite{koelemeij,leonard05}.
Molecular lines to the red of the $D_2$ atomic transition in helium dimers associated with
the He$(2\,^3\!S)$--He$(2\,^3\!P_2)$ asymptote are of interest for control of the scattering
length using an ``optical Feshbach resonance''~\cite{koelemeij,kim05}.
Published data are scarce on the long-range part of the He$(2\,^3\!S)$--He$(2\,^3\!P)$
potential energies, which determine the energy level structures of the ultralong-range
dimers formed in photoassociation of ultracold metastable helium atoms.


In this paper, we present perturbation theoretic calculations of the coefficients
determining the potential energies at large internuclear separations using accurate
variational wave functions in Hylleraas coordinates.  We evaluate the coefficient $C_{3}$ of
the first order resonant dipole-dipole energy, and the van der Waals coefficients $C_{6}$,
$C_{8}$, and $C_{10}$ of the second order energies arising from the mutual perturbations of
instantaneous electric dipole, quadrupole, and octupole interactions.  We also evaluate for
the third order energy the coefficient $C_{9}$ of the leading term.  We establish definitive
values including treatment of the finite nuclear mass for the
${}^3$He($2\,^3\!S$)--${}^3$He($2\,^3\!P$) and ${}^4$He($2\,^3\!S$)--${}^4$He($2\,^3\!P$)
interactions. Definitive values for the $C_6$, $C_8$, $C_{9}$, and $C_{10}$ coefficients are
established.

In this work,  atomic units are used throughout. For the He$(2\,^3\!S)$--He$(2\,^3\!P)$
system, the zeroth-order wave function appropriate for the perturbation calculation of the
long-range interaction can be written in the form~\cite{yan96}:
\begin{eqnarray}
\Psi^{(0)} (M,\pm) &=& \frac{1}{\sqrt 2}[\Psi_{n_{a}}(\mbox{\boldmath{$\sigma$}}) \Psi_{n_b}
(1 M;\rhoo) \pm \Psi_{n_{a}}(\rhoo)\Psi_{n_{b}}(1 M;\mbox{\boldmath{$\sigma$}})]\,, \label
{eq:ap1}
\end{eqnarray}
where $\Psi_{n_{a}}$ is the metastable He$(2\,^3\!S)$ wave function, $\Psi_{n_{b}}$ is the
wave function for the He$(2\,^3\!P)$ atom with magnetic quantum number $M$,
$\mbox{\boldmath{$\sigma$}}$ and $\rhoo$ are the collection of coordinates of the two helium
atoms in the laboratory reference frame, and the $``\pm"$ sign indicates the gerade and
ungerade states. The corresponding zeroth-order energy is
$E_{n_{a}n_{b}}^{(0)}=E^{(0)}_{n_{a}}+E^{(0)}_{n_{b}}$. At large internuclear distances, the
interaction potential between the two helium atoms can be expanded as an infinite series in
powers of $1/R$ \cite{yan96}
\begin{eqnarray}
V &=& \sum_{\ell=0}^{\infty}\sum_{L=0}^{\infty} \frac{V_{\ell L}}{R^{\ell+L+1}}\,, \label
{eq:ap2}
\end{eqnarray}
where
\begin{equation}
V_{\ell L}  = 4\pi (-1)^L (\ell,L)^{-1/2} \sum_{\mu} K_{\ell L}^{\mu} \;
T^{(\ell)}_{\mu}(\mbox{\boldmath{$\sigma$}})\; T^{(L)}_{-\mu}(\rhoo) \,. \label {eq:ap3}
\end{equation}
$T^{(\ell)}_{\mu}(\mbox{\boldmath{$\sigma$}})$ and $T^{(L)}_{-\mu}(\rhoo)$  are the atomic
multipole tensor operators defined as
\begin{equation}
T^{(\ell)}_{\mu}(\mbox{\boldmath{$\sigma$}}) = \sum_{i} Q_i \sigma_i^\ell Y_{\ell \mu}(\hat
{\mbox{\boldmath{$\sigma$}}}_i)\, ,\label{mpol}
\end{equation}
and
\begin{equation}
T^{(L)}_{-\mu}(\rhoo) = \sum_{j} q_j \rho_j^L Y_{L -\mu}(\hat {\rhoo}_j)\, ,
\end{equation}
where $Q_i$ and $q_j$ are the charges on particles $i$ and $j$. The coefficient $K_{\ell
L}^{\mu}$ is
\begin{eqnarray}
K_{\ell L}^{\mu} &=& \left[ { {\ell+L}\choose {\ell+\mu} } { {\ell+L}\choose {L+\mu}
}\right]^{1/2}\, \label {eq:ap4}
\end{eqnarray}
and $(\ell,L,\ldots)=(2\ell+1)(2L+1)\ldots$ . According to the perturbation theory, the
first-order energy~\cite{yan96} can be written as
\begin{eqnarray}
V^{(1)} &=& -\frac{C_{3}({M,\pm})}{R^{3}}\,, \label{eq:ap5}
\end{eqnarray}
where
\begin{eqnarray}
C_{3}({0,\pm}) &=& \pm \frac{8\pi}{9} |\langle\Psi_{n_{a}}
(\mbox{\boldmath{$\sigma$}})||\sum_i Q_i \sigma_iY_{1}
(\hat{\mbox{\boldmath{$\sigma$}}}_i)||\Psi_{n_b}(1;\mbox{\boldmath{$\sigma$}})\rangle|^2 \,,
\label{eq:ap6}
\end{eqnarray}
and
\begin{eqnarray}
C_{3}({\pm 1,\pm}) &=& \mp \frac{4\pi}{9} |\langle\Psi_{n_{a}}
(\mbox{\boldmath{$\sigma$}})||\sum_i Q_i \sigma_iY_{1}
(\hat{\mbox{\boldmath{$\sigma$}}}_i)||\Psi_{n_b}(1;\mbox{\boldmath{$\sigma$}})\rangle|^2 \,.
\label{eq:ap7}
\end{eqnarray}
The second-order energy is
\begin{eqnarray}
V^{(2)} &=& -{\sum_{n_sn_t}}'\sum_{L_sM_s}\sum_{L_tM_t} \frac{ |\langle\Psi^{(0)}|V|
\chi_{n_s} (L_sM_s;\mbox{\boldmath{$\sigma$}})\omega_{n_t} (L_tM_t;{\rhoo})\rangle|^2 }
{E_{n_sn_t}-E^{(0)}_{n_an_b}}\,,\ \label{eq:ap8}
\end{eqnarray}
where $ \chi_{n_s} (L_sM_s;\mbox{\boldmath{$\sigma$}})\omega_{n_t} (L_tM_t;{\rhoo})$
 is one of the intermediate states with the energy eigenvalue
$E_{n_sn_t}=E_{n_s}^{(0)}+E_{n_t}^{(0)}$, and the prime in the summation indicates that the
terms with $E_{n_sn_t}=E^{(0)}_{n_an_b}$ should be excluded. Substituting (\ref {eq:ap2})
into (\ref {eq:ap8}), we obtain
\begin{eqnarray}
V^{(2)} &=& -{\sum_{n_sn_t}}'\sum_{L_sM_s}\sum_{L_tM_t} \frac{B_1\pm B_2 }
{E_{n_sn_t}-E^{(0)}_{n_an_b}}\,, \label{eq:ap9}
\end{eqnarray}
with
\begin{eqnarray}
B_1 &=& |\langle \Psi_{n_{a}}(\mbox{\boldmath{$\sigma$}}) \Psi_{n_{b}}(1 M;{\rhoo})|V|
\chi_{n_s} (L_sM_s;\mbox{\boldmath{$\sigma$}}) \omega_{n_t} (L_tM_t;{\rhoo}) \rangle|^2 \,,
\label {eq:ap10}
\end{eqnarray}
and
\begin{eqnarray}
B_2 &=& \langle \Psi_{n_{a}}(\mbox{\boldmath{$\sigma$}}) \Psi_{n_{b}}(1 M;{\rhoo})|V|
\chi_{n_s} (L_sM_s;\mbox{\boldmath{$\sigma$}}) \omega_{n_t} (L_tM_t;{\rhoo})
\rangle\nonumber\\
&\times& \langle \Psi_{n_{a}}({\rhoo}) \Psi_{n_{b}}(1 M;\mbox{\boldmath{$\sigma$}})|V|
\chi_{n_s} (L_sM_s;\mbox{\boldmath{$\sigma$}}) \omega_{n_t} (L_tM_t;{\rhoo}) \rangle\,.
\label {eq:ap11}
\end{eqnarray}
$B_2$ in Eq.(\ref{eq:ap11}) is the exchange interaction of the two states He($2\,^3\!S$) and
He($2\,^3\!P$). After applying the Wigner-Eckart theorem, we have
\begin{eqnarray}
{\sum_{n_sn_t}}'\sum_{L_sM_s}\sum_{L_tM_t}\frac{B_1}{E_{n_sn_t}-E^{(0)}_{n_an_b}} &=&
\sum_{LL'L_sL_t}\frac{C_{1}(L,L',L_s,L_t,M)}{R^{2L_s+L+L'+2}}\,, \label {eq:ap12}
\end{eqnarray}
with
\begin{eqnarray}
C_{1}(L,L',L_s,L_t,M)&=& \frac{1}{2\pi}G'_1(L,L',L_s,L_t,M) F^{(1)}(L,L',L_s,L_t)\,. \label
{eq:ap13}
\end{eqnarray}
In Eq. (\ref {eq:ap13}), $G'_1$ is the angular-momentum part and $F^{(1)}$ is the oscillator
strength part. Their expressions are
\begin{eqnarray}
G'_1(L,L',L_s,L_t,M) &=&(-1)^{L+L'}(L,L')^{1/2}\sum_{M_sM_t}
K_{L_sL}^{-M_s}K_{L_sL'}^{-M_s}\nonumber\\
&\times& {\left(\matrix{1&L&L_t\cr{-M}&M_s&M_t\cr}\right)}
{\left(\matrix{1&L'&L_t\cr{-M}&M_s&M_t\cr}\right)}\,, \label {eq:ap14}
\end{eqnarray}
and
\begin{eqnarray}
F^{(1)}(L,L',L_s,L_t)=\frac{3\pi}{2}{\sum_{n_sn_t}}'\frac{\bar{g}_{n_s;n_{a}n_a}
(L_s,0,0,L_s,L_s)\bar{g}_{n_t;n_{b}n_{b}}(L_t,1,1,L,L') } {(\Delta E_{n_sn_a} +\Delta
E_{n_tn_{b}})|\Delta E_{n_sn_{a}}\Delta E_{n_tn_{b}}|}\,, \label {eq:ap15}
\end{eqnarray}
with
\begin{eqnarray}
  \bar{g}_{n_s;n_{a}n_{b}}(L_s,L_1,L_2,\ell,\ell') &=&
\frac{8\pi}{(\ell,\ell')} \frac{\sqrt{|\Delta E_{n_sn_a}\Delta
E_{n_sn_{b}}|}}{\sqrt{(2L_1+1)(2L_2+1)}}\nonumber\\
&\times& \langle\Psi_{n_{a}}(L_1;\mbox{\boldmath{$\sigma$}}) ||\sum_i Q_i \sigma_i^{\ell}
Y_{\ell}(\hat{\mbox{\boldmath{$\sigma$}}}_i)||\chi_{n_s}
(L_s;\mbox{\boldmath{$\sigma$}})\rangle\nonumber\\
&\times&\langle\Psi_{n_{b}}(L_2;\mbox{\boldmath{$\sigma$}}) ||\sum_i Q_i \sigma_i^{\ell'}
Y_{\ell'}(\hat{\mbox{\boldmath{$\sigma$}}}_i)||\chi_{n_s}
(L_s;\mbox{\boldmath{$\sigma$}})\rangle\,,
 \label{eq:ap16}
\end{eqnarray}
and $\Delta E_{n_sn_a}=E^{(0)}_{n_s}-E^{(0)}_{n_{a}}$, {\it etc}. For the special case when
the two initial states $\Psi_{n_a}$ and $\Psi_{n_b}$ are the same and $\ell=\ell'$,
$\bar{g}_{n_s;n_{a}n_{a}}$ reduces to the absolute value of the $2^\ell$-pole oscillator
strength
\begin{eqnarray}
\bar{f}_{n_sn_{a}}^{\ell} &=& \frac{8\pi\Delta
E_{n_sn_a}}{(2\ell+1)^2(2L_1+1)}|\langle\Psi_{n_{a}}(L_1;\mbox{\boldmath{$\sigma$}})
||\sum_i Q_i \sigma_i^{\ell} Y_{\ell}(\hat{\mbox{\boldmath{$\sigma$}}}_i)||\chi_{n_s}
(L_s;{\mbox{\boldmath{$\sigma$}}})\rangle|^2\,.
 \label{eq:ap17}
\end{eqnarray}
Similarly, we have
\begin{eqnarray}
{\sum_{n_sn_t}}'\sum_{L_sM_s}\sum_{L_tM_t}\frac{B_2}{E_{n_sn_t}-E^{(0)}_{n_an_b}} &=&
\sum_{LL'L_sL_t} \frac{C_{2}(L,L',L_s,L_t,M)}{R^{L_s+L_t+L+L'+2}}\,, \label {eq:ab18}
\end{eqnarray}
\begin{eqnarray}
C_{2}(L,L',L_s,L_t,M)&=& \frac{1}{2\pi}G'_2(L,L',L_s,L_t,M) F^{(2)}(L,L',L_s,L_t)\,, \label
{eq:app19}
\end{eqnarray}
with
\begin{eqnarray}
G'_{2}(L,L',L_s,L_t,M)&=&(-1)^{L+Ls}(L,L')^{1/2}
\sum_{M_sM_t}(-1)^{M_s+M_t}K_{L_sL}^{-M_s}K_{L'L_t}^{M_t}\nonumber\\
&\times& {\left(\matrix{1&L&L_t\cr{-M}&M_s&M_t\cr}\right)}
{\left(\matrix{1&L'&L_s\cr{-M}&M_t&M_s\cr}\right)}\,, \label {eq:ap20}
\end{eqnarray}
\begin{eqnarray}
F^{(2)}(L,L',L_s,L_t)=\frac{3\pi}{2}{\sum_{n_sn_t}}'
\frac{\bar{g}_{n_s;n_{a}n_b}(L_s,0,1,L_s,L')\bar{g}_{n_s;n_{a}n_{b}}(L_t,0,1,L_t,L) }
{(\Delta E_{n_sn_a} +\Delta E_{n_tn_b})\sqrt{|\Delta E_{n_sn_a}\Delta E_{n_sn_b}\Delta
E_{n_tn_a}\Delta E_{n_tn_b}}|}\,. \label {eq:ap21}
\end{eqnarray}
Finally, the second-order energy is
\begin{eqnarray}
V^{(2)} &=&-\sum_{n\geq 3} \frac{C_{2n}({M,\pm})}{R^{2n}}\,, \label {eq:ap22}
\end{eqnarray}
where $C_{2n}({M,\pm})$ are the dispersion coefficients
\begin{eqnarray} C_{2n}({M,\pm})=\mathop{\sum_{LL'L_sL_t}}_{L+L'+2L_s+2=2n}
C_{1}(L,L',L_s,L_t,M)\pm \mathop{\sum_{LL'L_sL_t}}_{L+L'+L_s+L_t+2=2n}
  C_{2}(L,L',L_s,L_t,M)\,. \label {eq:ap23}
\end{eqnarray}

According to third-order perturbation theory, the third-order energy correction is
\begin{eqnarray}
V^{(3)} &=&{\sum_{n_u
 n_v}}'{\sum_{n_sn_t}}'\sum_{L_sM_s}\sum_{L_uM_u}\sum_{L_vM_v}\sum_{L_tM_t}
\frac{{D_1}}
 {(E_{n_sn_t}-E^{(0)}_{n_an_b})(E_{n_un_v}-E^{(0)}_{n_an_b})}\nonumber\\
 &+& {\sum_{n_sn_t}}'\sum_{L_sM_s}\sum_{L_tM_t} \frac{ D_2 }
{(E_{n_sn_t}-E^{(0)}_{n_an_b})^2}
  \,,\
 \label{eq:tap8}
 \end{eqnarray}
where $D_1$ and $D_2$ are
\begin{eqnarray}
D_1 &=&  \langle\Psi^{(0)}|V| \chi_{n_s} (L_sM_s;\mbox{\boldmath{$\sigma$}})\omega_{n_t}
(L_tM_t;{\rhoo})\rangle \langle \chi_{n_s} (L_sM_s;\mbox{\boldmath{$\sigma$}})\omega_{n_t}
(L_tM_t;{\rhoo})|V| \chi_{n_u} (L_uM_u;\mbox{\boldmath{$\sigma$}}) \nonumber\\
& & \omega_{n_v} (L_vM_v;{\rhoo})\rangle \langle\chi_{n_u}
(L_uM_u;\mbox{\boldmath{$\sigma$}})\omega_{n_v} (L_vM_v;{\rhoo})|V| \Psi^{(0)}\rangle \,,\
\label{eq:tap9}
\end{eqnarray}
\begin{eqnarray}
D_2 &=&  -\langle\Psi^{(0)}|V|\Psi^{(0)}\rangle |\langle\Psi^{(0)}|V| \chi_{n_s}
(L_sM_s;\mbox{\boldmath{$\sigma$}})\omega_{n_t} (L_tM_t;{\rhoo})\rangle|^2\,.\
\label{eq:tap10}
\end{eqnarray}
For the He$(2\,^3\!S)$--He$(2\,^3\!P)$ system, following a procedure similar to the
second-order perturbation, the third-order energy correction can be expanded in terms of
powers of $1/R$:
\begin{eqnarray}
V^{(3)} &=&-\frac{C_9(M,\pm)}{R^9}-\frac{C_{11}(M,\pm)}{R^{11}}-\cdots.
\end{eqnarray}
In this work, we do not present a complete third order calculation. We only consider the
leading term $C_9(M,\pm)$ of $V^{(3)}$, which is comparable to the smallest term of
$V^{(2)}$.
\begin{eqnarray}
 C_9(M,\pm)&=&C^{D_1}_9(M,\pm)+C^{D_2}_9(M,\pm),
\end{eqnarray}
where
\begin{eqnarray}
 &&C^{D_1}_9(M,\pm)= {\sum_{n_u
 n_v}}'{\sum_{n_sn_t}}'\sum_{L_tL_u}\frac{\mp (4\pi)^3 G_{D_1}(L_t,L_u,M)}
 {(E_{n_sn_t}-E^{(0)}_{n_an_b})(E_{n_un_v}-E^{(0)}_{n_an_b})}\times\nonumber\\
&& \langle\Psi_{n_{a}}(\mbox{\boldmath{$\sigma$}}) ||\sum_i Q_i \sigma_i
Y_{1}(\hat{\mbox{\boldmath{$\sigma$}}}_i)||\chi_{n_s} (1;\mbox{\boldmath{$\sigma$}})\rangle
\langle\Psi_{n_{b}}(1;{\rhoo}) ||\sum_j
Q_j {\rho}_jY_{1}(\hat{\rhoo}_i)||\omega_{n_t} (L_t;{\rhoo})\rangle \times\nonumber\\
&& \langle\chi_{n_s}(1;\mbox{\boldmath{$\sigma$}}) ||\sum_i Q_i \sigma_i
Y_{1}(\hat{\mbox{\boldmath{$\sigma$}}}_i)||\chi_{n_u}
(L_u;\mbox{\boldmath{$\sigma$}})\rangle \langle\omega_{n_t}(L_t;{\rhoo}) ||\sum_j Q_j
{\rho}_jY_{1}(\hat{\rhoo}_i)||\omega_{n_v} (1;{\rhoo})\rangle  \times\nonumber\\
 &&  \langle\chi_{n_u}(L_u;\mbox{\boldmath{$\sigma$}}) ||\sum_i Q_i \sigma_i
Y_{1}(\hat{\mbox{\boldmath{$\sigma$}}}_i)||\Psi_{n_{b}}
(1;\mbox{\boldmath{$\sigma$}})\rangle \langle\omega_{n_v}(1;{\rhoo}) ||\sum_j Q_j
{\rho}_jY_{1}(\hat{\rhoo}_i)||\Psi_{n_{a}} ({\rhoo})\rangle\,, \ \label{eq:tap11}
\end{eqnarray}
\begin{eqnarray}
G_{D_1}(L_t,L_u,M)&=&\sum_{M_tM_uM_z}
\frac{(-1)^{M_z+L_t+L_u-M_u}}{81}K_{1\,1}^{-M_z}K_{1\,1}^{M_z-M_u}K_{1\,1}^{M_u-M}
\nonumber\\
&\times& {\left(\matrix{1&1&L_t\cr{-M}&M_z&M_t\cr}\right)}
{\left(\matrix{1&1&L_u\cr{-M_z}&M_z-M_u&M_u\cr}\right)}\nonumber\\
&\times& {\left(\matrix{L_t&1&1\cr{-M_t}&M_u-M_z&M-M_u\cr}\right)}
{\left(\matrix{L_u&1&1\cr{-M_u}&M_u-M&M\cr}\right)}
\end{eqnarray}
\begin{eqnarray}
 C^{D_2}_9(M,\pm) &=& \frac{-C_3(M,\pm)}{2\pi}\sum_{L_t} G'_{1}(1,1,1,L_t,M)
F^{D_2}(L_t)
 \,, \ \
\end{eqnarray}
\begin{eqnarray}
F^{D_2}(L_t)=\frac{3\pi}{2}{\sum_{n_sn_t}}'\frac{\bar{g}_{n_s;n_{a}n_a}
(1,0,0,1,1)\bar{g}_{n_t;n_{b}n_{b}}(L_t,1,1,1,1) } {(\Delta E_{n_sn_a} +\Delta
E_{n_tn_{b}})^2|\Delta E_{n_sn_{a}}\Delta E_{n_tn_{b}}|}\,.
\end{eqnarray}

The nonrelativistic Hamiltonian for a helium atom in a laboratory frame is
\begin{eqnarray}
H &=& -\frac{1}{2m_n}\nabla_{R_n}^2
-\frac{1}{2m_e}\nabla_{R_1}^2-\frac{1}{2m_e}\nabla_{R_2}^2 + \frac{q_nq_e}{|{\bf R}_n-{\bf
R}_1|} +\frac{q_nq_e}{|{\bf R}_n-{\bf R}_2|}+\frac{q_e^2}{|{\bf R}_1-{\bf R}_2|}\,, \label
{eq:t16}
\end{eqnarray}
where $m_n$ and $m_e$ are the masses of nucleus and electron respectively, $q_n$ and $q_e$
are their charges, and ${\bf R}_n$, ${\bf R}_1$, and ${\bf R}_2$ are the corresponding
position vectors relative to the laboratory frame. Introduce the following transformation
\begin{eqnarray}
{\bf r}_1 &=& {\bf R}_1 - {\bf R}_n\,, \\
{\bf r}_2 &=& {\bf R}_2 - {\bf R}_n\,,  \\
{\bf X} &=& \frac{1}{M_T} (m_n{\bf R}_n+m_e {\bf R}_1+m_e {\bf R}_2)\,, \label {eq:t17}
\end{eqnarray}
where $M_T=m_n+2m_e$ is the total mass of helium atom and ${\bf X}$ is the position vector
for the center of mass of helum. In the center-of-mass frame, the Hamiltonian becomes
\begin{eqnarray}
H &=& - \frac{1}{2\mu_e}\nabla_{{\bf r}_1}^2 - \frac{1}{2\mu_e}\nabla_{{\bf r}_2}^2
-\frac{1}{m_n}\nabla_{{\bf r}_1}\cdot\nabla_{{\bf r}_2}
+\frac{q_nq_e}{r_1}+\frac{q_nq_e}{r_2} +\frac{q_e^2}{r_{12}}\,, \label {eq:t20}
\end{eqnarray}
where $r_{12}=|{\bf r}_1-{\bf r}_2|$ and $\mu_e=m_em_n/(m_e+m_n)$ is the reduced mass
between the electron and the nucleus. The eigenvalue spectrum and corresponding
eigenfunctions are obtained by diagonalizing the Hamiltonian (\ref{eq:t20}) in the
correlated Hylleraas basis set~\cite{yan96}
\begin{eqnarray}
 \{r_1^i\,r_2^j\,r_{12}^k\,e^{-\alpha r_1-\beta r_2}
{\cal Y}_{{\ell_1}{\ell_2}}^{{LM}}({\hat{\bf r}}_1,{\hat{\bf r}}_2)\} \,,
\end{eqnarray}
where ${\cal Y}_{{\ell_1}{\ell_2}}^{{LM}}({\hat{\bf r}}_1,{\hat{\bf r}}_2)$ is the coupled
spherical harmonics for the two electrons forming a common eigenstate of ${\bf L}^2$ and
$L_z$. Except for some truncations, all terms are included in the basis such that
\begin{eqnarray}
i+j+k\leq\Omega \,,
\end{eqnarray}
with $\Omega$ being an integer. As $\Omega$ increases, the size of basis set is increased
progressively.

It is necessary to transform the transition operator
\begin{eqnarray}
T_{\ell} &=&  q_n R_n^\ell Y_{\ell 0}({\hat{\bf R}}_n)+q_e R_1^\ell Y_{\ell 0}({\hat{\bf
R}}_1)+q_e R_2^\ell Y_{\ell 0}({\hat{\bf R}}_2)\, \label {eq:ap54}
\end{eqnarray}
into the center-of-mass coordinates~\cite {zhy04}. The transformed $T_\ell$ with $\ell=1$,
2, and 3 for the dipole, quadrupole, and octupole moments for a neutral helium are
\begin{eqnarray}
&&T_1 = -\sum_{j=1}^2r_j Y_{10}({\hat{\bf r}}_j)\,,\\
&&T_2 = -\bigg(1-2\frac{m_e}{M_T}\bigg)\sum_{j=1}^2r_j^2 Y_{20}({\hat{\bf r}}_j)
+\sqrt{\frac{30}{\pi}}\frac{m_e}{M_T}\, r_1r_2
(\hat{\bf r}_1\otimes \hat{\bf r}_2)^{(2)}_0\,,\\
&&T_3 = -\bigg[1-3\frac{m_e}{M_T}+3\bigg(\frac{m_e}{M_T}\bigg)^2\bigg]
\sum_{j=1}^2r_j^3 Y_{30}({\hat{\bf r}}_j)\nonumber\\
&&+\frac{3}{2}\sqrt{\frac{35}{2\pi}} \bigg[\frac{m_e}{M_T}-3
\bigg(\frac{m_e}{M_T}\bigg)^2\bigg] \bigg[r_1^2r_2 ((\hat{\bf r}_1\otimes \hat{\bf
r}_1)^{(2)}\otimes \hat{\bf r}_2)^{(3)}_0
     +r_2^2r_1 ((\hat{\bf r}_2\otimes \hat{\bf r}_2)^{(2)}\otimes \hat{\bf
r}_1)^{(3)}_0 \bigg]\,. \label {eq:ap54_1}
\end{eqnarray}
It is noted that for the case of infinite nuclear mass, the above operators reduce to
\begin{eqnarray}
T^\infty_\ell = -\sum_{j=1}^2r_j^\ell Y_{\ell 0}({\hat{\bf r}}_j)\,. \label {eq:ap54_2}
\end{eqnarray}
For the finite nuclear mass case, however, $T_\ell$ can not be obtained by a simple mass
scaling from $T^\infty_\ell$, except $T_1$ which does not contain $m_e/M_T$ explicitly for a
neutral system.

Table~\ref{g1} gives the convergence pattern of $C_{6}({M,\pm})$ for the case of infinite
nuclear mass as the sizes of basis sets, including the two initial states and the four
intermediate states, increase progressively. Table~\ref{g2} lists the contributions to
$C_{6}({M,\pm})$ from three pairs of intermediate symmetries $(\,^3\!P,\,^3\!S)$,
$(\,^3\!P,(pp)\,^3\!P)$ doubly-excited states, and $(\,^3\!P,\,^3\!D)$. Our final results
for $C_{3}({M,\pm})$, $C_{6}({M,\pm})$, $C_{8}({M,\pm})$, $C_{9}({M,\pm})$, and
$C_{10}({M,\pm})$ are presented in Table~\ref{g3}. To our knowledge, no definitive
calculations have been reported for the dispersion coefficients $C_{6}({M,\pm})$,
$C_{8}({M,\pm})$, $C_{9}({M,\pm})$, and $C_{10}({M,\pm})$. Table~\ref{g4} is a comparison
with the existing values of $C_{3}({M,\pm})$ and $C_{6}({M,\pm})$. Our values for
$C_{3}({M,\pm})$ differ from those given in Refs.~\cite{cohen2} and \cite{venleo}. The
origins of these discrepancies are uncertain. Drake~\cite{drake} tabulated the oscillator
strengths of helium, including $^3\!\,$He and $^4\!\,$He. In order to extract the square of
the transition matrix element connecting $2\,^3\!S$ and $2\,^3\!P$, besides some numerical
constants, the tabulated value should be multipied by $1+2m_e/m_n$ and divided by the
transition energy corrected for the finite nuclear mass. Our results are in perfect
agreement with Drake's values such obtained.  For $C_{6}({M,\pm})$, our values differ from
the values used in the work of Venturi {\it et al.}~\cite{venleo} at the level of 0.8\% and
0.9\% respectively for $C_{6}({0,\pm})$ and $C_{6}({\pm 1,\pm})$.

\acknowledgments We are grateful to Dr. Jim Mitroy for his pointing out dispersion
coefficients $C_{9}({M,\pm})$ from the third-order energy correction. This work is supported
by the Natural Sciences and Engineering Research Council of Canada, by the ACRL of the
University of New Brunswick, by the SHARCnet, and by NSF through a grant for the Institute
of Theoretical Atomic, Molecular and Optical Physics (ITAMP) at Harvard University and
Smithsonian Astrophysical Observatory.

\newpage
\begin{longtable}{c c c c c c c c}
\caption{\label{g1} Convergence characteristics of $C_{6}({M,\pm})$, in atomic units, for
the $^\infty\!\,$He($2\,^3\!S$)--$^\infty\!\,$He($2\,^3\!P$) system. $N_{\,^3\!S}$,
$N_{\,^3\!P}$, $N^*_{\,^3\!S}$, $N^*_{\,^3\!P}$, $N_{(pp)\,^3\!P}$, and $N_{\,^3\!D}$ denote
respectively the sizes of bases for the two initial states and the four intermediate states
of symmetries $\,^3\!S$, $\,^3\!P$, $(pp)\,^3\!P$, and $\,^3\!D$.
}\\
\hline\hline \multicolumn{1}{c}{$N_{\,^3\!S}$}& \multicolumn{1}{c}{$N_{\,^3\!P}$}&
\multicolumn{1}{c}{$N^*_{\,^3\!S}$}& \multicolumn{1}{c}{$N^*_{\,^3\!P}$}&
\multicolumn{1}{c}{$N_{(pp)\,^3\!P}$}& \multicolumn{1}{c}{$N_{\,^3\!D}$}&
\multicolumn{1}{c}{$C_{6}({0,\pm})$}&
\multicolumn{1}{c}{$C_{6}({\pm 1,\pm})$}\\
\hline
1330    &1360 &560 &1360 & 1230 &853 &2640.233\,681 &1862.572\,368     \\
1540    &1632 &680 &1632 & 1430 &1071 &2640.233\,694 &1862.572\,376     \\
1771    &1938 &816 &1938 & 1650 &1323 &2640.233\,700 &1862.572\,380       \\
\hline\hline
\end{longtable}
\begin{longtable}{l l l}
\caption{\label{g2} Contributions to $C_{6}({M,\pm})$, in atomic units, for the
$^\infty\!\,$He($2\,^3\!S$)--$^\infty\!\,$He($2\,^3\!P$) system from $(\,^3\!P,\,^3\!S)$,
$(\,^3\!P,(pp)\,^3\!P)$, and $(\,^3\!P,\,^3\!D)$ symmetries.
}\\
\hline\hline \multicolumn{1}{l}{Symmetries}& \multicolumn{1}{c}{$C_{6}({0,\pm})$}&
\multicolumn{1}{c}{$C_{6}({\pm 1,\pm})$}\\
\hline
$(\,^3\!P,\,^3\!S)$ &684.091969(2) &171.0229919(1)      \\
$(\,^3\!P,(pp)\,^3\!P)$ &1.31648932(2) &3.29122329(3)      \\
$(\,^3\!P,\,^3\!D)$ &1954.825248(4) & 1688.258168(3)      \\
\hline\hline
\end{longtable}
\newpage
\begin{longtable}{l l l l}
\caption{\label{g3} $C_{3}({M,\pm})$, $C_{6}({M,\pm})$, $C_{8}({M,\pm})$, $C_{9}({M,\pm})$,
and $C_{10}({M,\pm})$, in atomic units, for the He($2\,^3\!S$)--He($2\,^3\!P$) system.
}\\
\hline\hline \multicolumn{1}{l}{Mass}& \multicolumn{1}{c}{$^\infty\!\,$He--$^\infty\!\,$He}&
\multicolumn{1}{c}{$^4\!\,$He--$^4\!\,$He}& \multicolumn{1}{c}{$^3\!\,$He--$^3\!\,$He}\\
\hline
  $C_{3}({0,\pm})$ &$\pm12.8154931075(4)$ &$\pm12.8181751205(4)$
&$\pm12.8190526019(4)$\\
  $C_{3}({\pm 1,\pm})$ &$\mp6.4077465536(2)$
&$\mp6.4090875603(2)$&$\mp6.4095263011(2)$\\
$C_{6}({0,\pm})$  &2640.2338(1)&2641.5083(2)&2641.9255(3)\\
$C_{6}({\pm 1,\pm})$ &1862.5724(1)&1863.4726(2)&1863.7674(4)\\
$C_{8}({0,+})$&311901.2(4)&311955.4(5)&311972.8(1)\\
$C_{8}({0,-})$&1541993(2)&1542352(1)&1542470(1)\\
 $C_{8}({\pm 1,+})$&168906.5(4)&168921.6(3)&168926.5(2)\\
$C_{8}({\pm 1,-})$&103017.3(3)&103039.5(4)&103046.7(3)\\
$C_{9}({0,\pm})$ &$\pm512059.227(6)$ &$\pm512572.343(6)$ &$\pm512740.318(6)$\\
  $C_{9}({\pm 1,\pm})$ &$\mp117073.536(2)$ &$\mp117199.211(2)$&$\mp117240.354(2)$\\
$C_{10}({0,+})$&$2.922482(3)\times 10^{7}$&$2.922304(5)\times
10^{7}$&$2.922244(3)\times10^{7}$\\
$C_{10}({0,-})$&$1.857456(3)\times 10^{8}$&$1.8574503(3)\times 10^{8}$&$1.8574492(3)\times
10^{8}$\\
$C_{10}({\pm 1,+})$&$1.611301(3)\times 10^{7}$&$1.611325(3)\times 10^{7}$&$1.611334(5)\times
10^{7}$\\
$C_{10}({\pm 1,-})$&$2.40597(3)\times 10^{6}$&$2.40608(1)\times 10^{6}$&$2.40613(2)\times
10^{6}$ \\
\hline\hline
\end{longtable}
\begin{longtable}{l l l l l}
\caption{\label{g4} Comparison of $C_{3}({M,\pm})$ and $C_{6}({M,\pm})$ for the
$^4\!\,$He($2\,^3\!S$)--$^4\!\,$He($2\,^3\!P$) system.
}\\
\hline\hline \multicolumn{1}{l}{Author}& \multicolumn{1}{c}{$C_{3}({0,\pm})$}&
\multicolumn{1}{c}{$C_{3}({\pm 1,\pm})$}& \multicolumn{1}{c}{$C_{6}({0,\pm})$}&
\multicolumn{1}{c}{$C_{6}(\pm 1,\pm)$}\\
\hline
Venturi {\it et al.}~\cite{venleo} &$\pm12.82044$ &$\mp6.41022$ &2620.76 &1846.60\\
L\'{e}onard {\it et al.}~\cite{cohen2} &$\pm12.810(6)$ &$\mp6.405(3)$ & &\\
This work & $\pm12.8181751205(4)$
&$\mp6.4090875603(2)$&2641.5083(2)&1863.4726(2)\\
 \hline\hline
\end{longtable}

\end{document}